\documentclass[10pt,twocolumn,superscriptaddress,longbibliography,prx]{revtex4-2}

\usepackage{amsmath, bm, amsfonts, amssymb}     
\usepackage{graphicx}                       
\usepackage{xfrac}                          
\usepackage{grffile}                        
\usepackage{xcolor}

\newcommand{\sig}{\Sigma_0}
\newcommand{\gdot}{\dot{\gamma}}
\newcommand{\tw}{t_{\rm w}}
\newcommand{\tauf}{\tau_{\rm f}}
\newcommand{\tref}{\tau_{\rm ref}}
\newcommand{\taumin}{\tau_{\rm min}}
\newcommand{\gdotmin}{\dot{\gamma}_{\rm min}}

\usepackage[colorlinks, citecolor=blue, urlcolor=red, linkcolor=blue]{hyperref}

\begin{document}

\title{Power law creep and delayed failure of gels and fibrous materials under stress}
\author{Henry A. Lockwood}
\affiliation{Department of Physics, Durham University, Science Laboratories,
  South Road, Durham DH1 3LE, UK}
 \author{Molly H. Agar}
\affiliation{Department of Physics, Durham University, Science Laboratories,
  South Road, Durham DH1 3LE, UK}
\author{Suzanne M. Fielding}
\affiliation{Department of Physics, Durham University, Science Laboratories,
  South Road, Durham DH1 3LE, UK}

\begin{abstract}

Motivated by recent experiments studying the creep and breakup of a protein gel under stress, we introduce a simple mesoscopic model for the irreversible failure of gels and fibrous materials, and demonstrate it to capture much of the phenomenology seen experimentally. This includes a primary creep regime in which the shear rate decreases as a power law over several decades of time, a secondary crossover regime in which the shear rate attains a minimum, and a tertiary regime in which the shear rate increases dramatically up to a finite time singularity, signifying irreversible material failure.  The model also captures a linear Monkman-Grant scaling of the failure time  with the earlier time at which the shear rate  attained its minimum, and a Basquin-like power law scaling of the failure time with imposed stress, as seen experimentally. The model furthermore predicts a slow accumulation of low levels of material damage during primary creep, followed by the growth of fractures leading to sudden material failure, as seen experimentally.

\end{abstract}

\maketitle

\section{Introduction}

Materials subject to loading sustained over a lengthy duration will often catastrophically fail after a long induction time, without apparent warning. This has important implications in myriad contexts, including the shelf life of products, the integrity of construction materials, and the prediction of geophysical phenomena such as avalanches, mudslides and earthquakes. In the context of soft gels and fibrous network materials, delayed failure under constant load has been observed in transient 
gels~\cite{poon1999delayed,skrzeszewska2010fracture}, polymeric gels~\cite{bonn1998delayed}, collagen networks~\cite{gobeaux2010power}, protein gels~\cite{leocmach2014creep}, colloidal gels~\cite{gopalakrishnan2007delayed,gibaud2010heterogeneous,grenard2014timescales,sprakel2011stress,cho2022yield,aime2018microscopic,moghimi2021yielding,landrum2016delayed}, hydrogels~\cite{karobi2016creep}, biopolymer gels~\cite{pommella2020role},  fibre composites~\cite{nechad2005creep} and paper~\cite{rosti2010fluctuations}. 
In some cases yielding  is reversible,
with the material reforming and recovering elasticity once the load is removed. In others, failure is irreversible and the original  network is permanently broken.

In this work, we focus on the irreversible failure of gels and fibrous materials, motivated in particular by recent experiments demonstrating that the behaviour of a protein gel under sustained loading is strikingly similar to that of brittle solids~\cite{leocmach2014creep,bauland2023non}. In these experiments, an initial regime of slow creep over several decades of time $t$ later gave way to catastrophic material failure at a time $\tauf$. In the primary creep regime, the shear strain $\gamma$ increased up to $O(1)$ in a sublinear way, $\gamma\sim t^{1-\alpha}$ with $\alpha<1$, with the  shear rate decreasing as $\gdot\sim t^{-\alpha}$, reminiscent of Andrade creep in solids~\cite{andrade1910viscous,miguel2002dislocation}. In a secondary crossover regime for times $0.1\lesssim t/\tauf \lesssim 0.9$, the shear rate attained a minimum $\gdotmin$ at time $\taumin$. The shear rate then increased dramatically as  $\gdot\sim (\tauf-t)^{-1}$ in a final tertiary regime, diverging in a finite time singularity at $\tauf$,  signifying catastrophic material failure. Across all three regimes, the shear rate fit the single master relation
\begin{equation}
\frac{\gdot(t)}{\gdotmin}=\lambda\left(\frac{t}{\tauf}\right)^{-\alpha}+\frac{\mu}{1-t/\tauf}.
\label{eqn:master}
\end{equation}

The same experiments demonstrated the failure time to decrease with imposed stress as $\tauf\sim\sig^{-\beta}$, reminiscent of the Basquin law of fatigue in solids~\cite{oh1910exponential}, and to scale linearly with $\taumin$ via a Monkman-Grant relation~\cite{monkman1956empirical}, $\tauf=\taumin/c$ with $c<1$. Importantly, this suggests that sudden failure at time $\tauf$ might be {\em forecast} once a shear rate minimum has earlier been observed at  $\taumin$. 

By combining rheology with optical and ultrasonic imaging~\cite{gallot2013ultrafast}, these experiments also tracked the strain field within the gel, in tandem with the global strain $\gamma$. In the primary creep regime, no macroscopic strain localisation was seen, although the authors noted that this  does not rule out local rearrangements on scales below those imaged. Indeed, separate recovery tests suggested low levels of irreversible material damage during primary creep. Fractures then developed during the secondary regime, and grew to cause failure in the tertiary regime. 

Motivated by these experiments, we introduce a simple mesoscopic model for the irreversible breakup of gels. Simulations in the step-stress protocol show it to capture the experimental phenomenology just described: (i) primary creep with  decreasing shear rate $\gdot\sim t^{-\alpha}$, (ii) a secondary cross-over regime in which the shear rate attains a minimum at time $\taumin$, (iii) a tertiary regime in which the shear rate increases dramatically as $(\tauf-t)^{-1}$, leading to (iv) catastrophic failure in a finite time singularity at time $\tauf$,  (v) a linear Monkman-Grant scaling of $\tauf$ with $\taumin$, (vi) a Basquin-like scaling of $\tauf$ with $\sig$, (vii) accumulating low levels of damage during primary creep, followed by (viii) the growth of macroscopic fractures leading to catastrophic material failure.

Although we shall mostly compare our results  with the particular experiments of Ref.~\cite{leocmach2014creep}, other experiments~\cite{bauland2023non} report the same phenomenology, with quantitative values of the exponents $\alpha,\beta$ however being non-universal across materials. Therefore, our focus will be on robustly capturing all the qualitative behaviour just  described, while also characterising how the  values of exponents such as $\alpha,\beta$, etc.  depend on model parameters.

\section{Model}

Our approach will be to model a macroscopic sample of gel by considering the underlying physics at the mescoscopic level of a few gel strands and their associated crosslinks. In the spirit of existing elastoplastic models~\cite{nicolas2018deformation}, each local collection of strands and crosslinks is represented by a single elastoplastic element, with local shear strain $l$ and stress $kl$ (discarding normal stresses). The macroscopic elastoplastic stress $\sigma$ is the average of these local elemental ones. Before any local failure event, each element affinely follows the macroscopic shear strain, $\dot{l}=\gdot$, giving elastic response. To model the strain-induced breakage of gel strands, when any element's elastic energy $E=\tfrac{1}{2}kl^2$ exceeds a local yielding threshold $E_{\rm y}$, it fails plastically with stochastic rate $\tau_0^{-1}$. Activated strand breakage is also captured via a yielding rate $\tau_0^{-1}\exp\left[(\tfrac{1}{2}kl^2-E_{\rm y})/x\right]$ for elements even below threshold. In materials for which the energy barriers to strand breakage are comparable with thermal energies, we take $x=k_{\rm B}T$. For much larger barriers, $x$ is taken to be a mean field temperature modeling strand breakage due to the propagation of stresses from breakages elsewhere. On the timescales of creep and yielding, the breakage of any strand is assumed irreversible, with an infinite reformation time in comparison,  $\tref=\infty$. The associated element carries no stress thereafter.

To model the much slower process of gelation during a ``waiting time'' $-\tw<t<0$ prior to the switch-on of stress at $t=0$, we must of course model the (re)formation of gel strands. During gelation, therefore, we include an additional model ingredient: that after any element locally yields, it reforms with local strain $l=0$ and a new energy threshold drawn from a distribution $\rho(E)\sim \exp(-E/x_{\rm g})$ in which the model parameter $x_{\rm g}$ is the glass transition temperature.  As an initial condition at $t=-\tw$, we assume a distribution of thresholds $\rho(E)$. For $x<x_{\rm g}$, the model captures ageing, with dynamics that become slower over time as the system gels. We take element reformation to be infinitely fast compared with this slow process of gelation, with $\tref=0$. This is oversimplified: a fuller description might consider a finite reformation time $\tref$ consistently during gelation, creep and yielding. However, our focus here is not to model gelation itself, but the subsequent creep and yielding of a gel, however formed, under stress.

During gelation, with immediate element reformation after breakage via an assumed $\tref=0$, compared with the much slower timescale of gelation, the model is the same as the soft glassy rheology model~\cite{sollich1997rheology}. During creep, the model differs from the SGR model in having irreversible element breakage via an assumed $\tref=\infty$, compared with the much faster timescale of yielding.

So far, we have described the model as though the strain rate $\gdot(t)$ were the same for all elements. In fact, to allow for strain localisation during material failure, we consider an array of $S$ streamlines  stacked across the flow gradient direction $y$ of a shear cell of gap width $L_y$, with $M$ elastoplastic elements on each streamline~\cite{fielding2009shear}. Imposing force balance then allows us to calculate the shear rate profile $\gdot(y,t)$, which becomes highly heterogeneous as the sample yields. We do this via two different algorithms. The first considers a solvent stress of small viscosity $\eta=0.05$, additive to the elastoplastic stress~\cite{barlow2020ductile}, giving a total stress $\Sigma=\sigma+\eta\gdot$. The second has $\eta=0$~\cite{fielding2009shear}. The  results shown are converged to the limit $\eta\to 0$, suited to the relevant experimental regime, $\eta\ll k\tau_0$. The $\eta=0$ algorithm was needed only to obtain the scaling of shear rate with time in the final tertiary regime, as the gel finally breaks completely. To seed the formation of strain bands, we initialise our creep simulations  with a small heterogeneity: $\tw\to \tw\left[1+\epsilon \cos(2\pi y/L_y)\right]$. Indeed, in physical practice, most rheometers seed slight heterogeneity due to device curvature. For a fuller discussion of other possible sources of heterogeneity that might seed band formation, see Ref.~\cite{moorcroft2014shear}.

We choose units in which the attempt time $\tau_0=1$, local modulus $k=1$ and gap width $L_y=1$. We also set $x_{\rm g}=1$, thereby setting typical local yield strains of order unity. Unless otherwise stated we use  $S=10$ streamlines  with $M=100,000$ elements per streamline, a 
perturbation amplitude $\epsilon=0.1$, and numerical time-step $dt=0.01$. Parameters explored are then the imposed stress $\sig$, the waiting time $\tw$, and the noise temperature $x$. 

\begin{figure}[!t]
\includegraphics[width=\columnwidth]{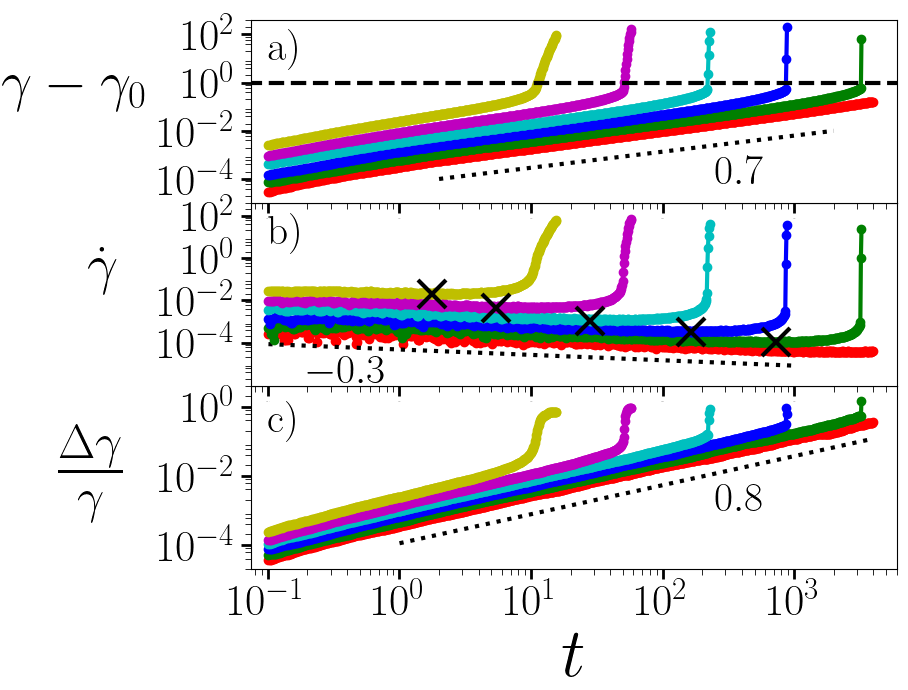}
\caption{Creep and failure under a shear stress imposed for times $t>0.0$. Stress $\sig =1.0, 1.2 \cdots 2.0$ in curves upwards. {\bf Top:} Strain $\gamma-\gamma_0$ as a function of time $t$ since the stress was imposed. Dashed line denotes the strain value $\gamma=1.0$ used to define the time $\tauf$ at which material failure occurs.  Dotted line shows the power $0.7=1-x$. Recall that $\gamma_0$ is the elastic strain  that arises immediately upon stress switch-on. {\bf Middle:} corresponding strain rate $\gdot(t)$. Dotted line shows the power $-0.3= -x$. Crosses indicate the minimum value of the strain rate $\gdotmin$ in each simulation, and the time $\taumin$ at which it occurs. {\bf Bottom:} corresponding growing strain heterogeneity (degree of shear banding), defined at any time $t$ by the standard deviation of strain across $y$, normalised by the global strain $\gamma(t)$. Dotted line shows the power $0.8$. Noise temperature $x=0.3$. Waiting time $t_w = 10^5$. $\eta=0.05$.
}
\label{fig:creepA}
\end{figure}

\section{Results}

Following the switch-on of a stress of magnitude $\sig$ at time $t=0$, we report the global shear strain $\gamma(t)$ (relative to the elastic strain $\gamma_0=\Sigma_0$ that arises immediately upon stress switch-on) and  shear rate $\gdot(t)$ in Fig.~\ref{fig:creepA}a,b).  The basic physics seen experimentally~\cite{leocmach2014creep} is clearly evident:  primary creep with increasing strain  $\gamma-\gamma_0\sim t^{1-\alpha}$ and decreasing strain rate $\gdot\sim t^{-\alpha}$ is followed by a sudden near-divergence of these quantities, signifying catastrophic failure. (True divergence is averted in Fig.~\ref{fig:creepA} via the solvent viscosity, which ensures a finite $\gdot=\sig/\eta$ even after all gel strands break.) The creep exponent  $\alpha=x$ for (noise) temperatures $x<1$. Experiments find $\alpha=0.85$ in casein gels~\cite{leocmach2014creep} and $\alpha=0.7$ in milk gels~\cite{bauland2023non}.

Alongside the global strain signal $\gamma(t)$, we also plot in Fig.~\ref{fig:creepA}c)  the degree to which the strain field  becomes heterogeneous, i.e., shear banded, as a function of the flow-gradient coordinate $y$ across the shear cell. Specifically, we characterise this degree of strain localisation, i.e., of shear banding, by defining the quantity $\Delta\gamma(t)$. This is the standard deviation in strain across the streamlines stacked across $y$ at any time $t$, normalised by the globally average strain $\gamma(t)$. This remains small during primary creep, consistent with the low levels of irreversible damage evidenced via strain recovery experiments~\cite{leocmach2014creep}.  The level of strain localisation  then increases dramatically as the material yields, as seen experimentally~\cite{leocmach2014creep}.

\begin{figure}[!t]
\includegraphics[width=\columnwidth]{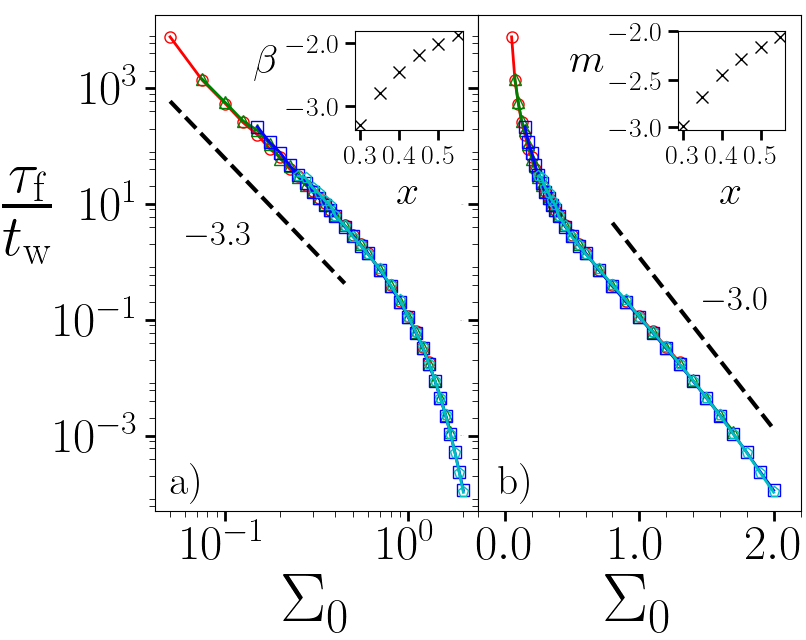}
\caption{Failure time $\tauf$ vs imposed stress $\sig$ at a noise temperature $x=0.3$. {\bf Left:} on a log-log scale, showing a power law dependence $\tauf\sim \sig^{-\beta}$ at low $\sig$. Dashed line shows power $\beta=3.3$. Inset: exponent $\beta$ vs. $x$. {\bf Right:} on a log-lin scale, showing an exponential dependence $\tauf\sim \exp(-m\sig)$ at larger $\sig$. Dashed line shows exponent $m=3.0$.  Inset: exponent $m$ vs. $x$. Waiting time $\tw=10^n$ with $n=3.0, 4.0, 5.0, 6.0$ for the red circles, green triangles, blue squares and cyan pentagons respectively. $\eta=0.05$.
}
\label{fig:tauf}
\end{figure}
\begin{figure}[!t]
\includegraphics[width=0.9\columnwidth]{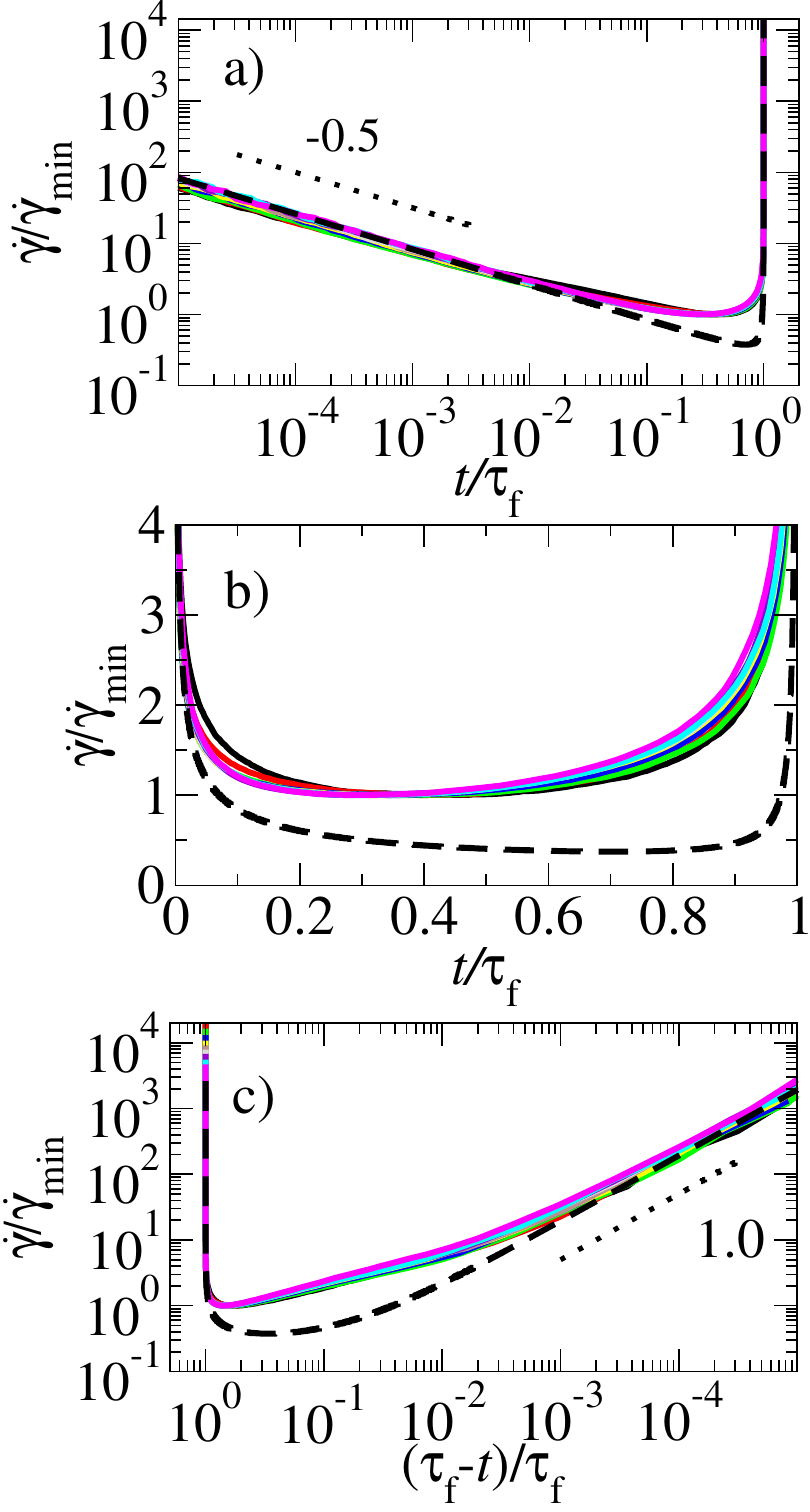}
\caption{Strain rate $\gdot$ normalised by its minimum value $\gdotmin$. {\bf Top:} Plotted versus time $t$ normalised by the failure time $\tauf$, with log-log axes to emphasise the primary creep regime, $\gdot\sim t^{-\alpha}$. Dotted line shows the power $\alpha=x=0.5$. {\bf Middle:} Plotted likewise, with lin-lin axes to emphasise the secondary crossover regime. {\bf Bottom:} plotted as a function of time remaining to failure $\tauf-t$, normalised by $\tauf$, to emphasise the tertiary regime of final failure. Note that the time axis is visually inverted so as to move towards failure from left to right. Dotted line shows the power $+1.0$. Dashed line in each panel shows the fit to the master scaling form of Eqn.~\ref{eqn:master} with  $\mu=0.019$ and $\lambda=0.26$. Imposed stress $\sig=0.1, 0.2\cdots 1.0$ in curves black, red, green, blue, yellow, brown, grey, violet, cyan and magenta. Noise temperature $x=0.5$, waiting time $\tw=10^9$. $M=10000$. $\eta=0.0$.}
\label{fig:creepB}
\end{figure}
\begin{figure}[!t]
\includegraphics[width=\columnwidth]{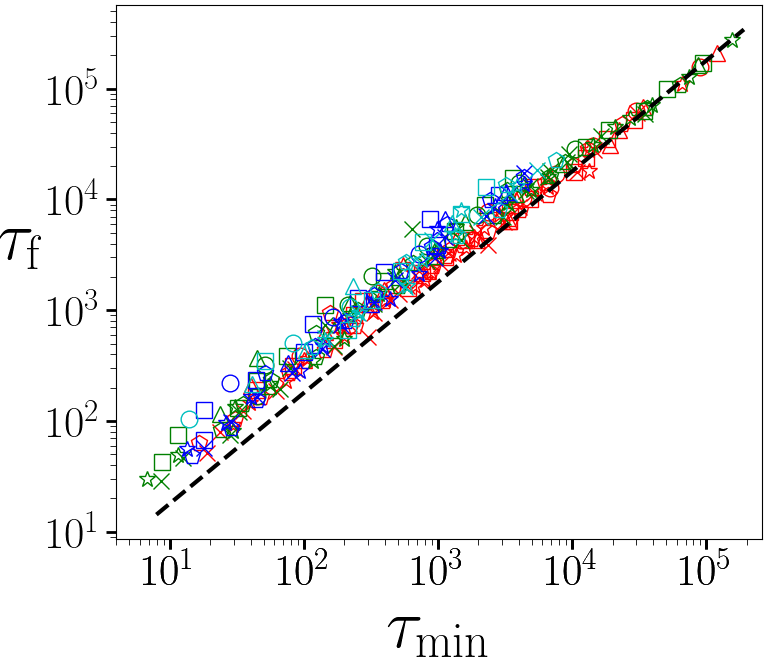}
\caption{Time $\tauf$ of final material failure versus the earlier time $\taumin$ at which the shear rate attained its minimum. Each data point corresponds to a single simulation. Data for noise temperatures $x=0.30, 0.35 \cdots 0.50$ are shown by  circles, triangles, squares, pentagons, stars and crosses respectively. Data for waiting times $\tw=10^n$ with $n=3.0, 4.0, 5.0, 6.0$ are shown by red, green, blue and cyan symbols respectively. Imposed stress $\sig=0.05,0.10,0.15\cdots 2.0$ increases right to left for each symbol shape and colour. Dashed line shows the linear Monkman-Grant relationship observed experimentally: $\tauf=\taumin/c$ with $c=0.56$. $\eta=0.05$.
}
\label{fig:MGrelation}
\end{figure}

As indicated by the dashed line,  we define the failure time $\tauf$  as the time at which the global strain attains a value equal to unity in Fig.~\ref{fig:creepA}a). (In any simulations with $\eta=0$,  we instead define $\tauf$ as the time at which the strain rate actually diverges. We have checked that these definitions give the same scaling of $\tauf$ with $\sig$ and $\taumin$, with only a small fractional quantitative difference.)
This is plotted vs the imposed stress $\sig$ in Fig.~\ref{fig:tauf}, for several values of the waiting time $\tw$ at a fixed noise temperature $x$. Collapse of the data when normalised as $\tauf/\tw$ indicates a scaling $\tauf\sim \tw$, consistent with the aging predicted by our model during gelation.  The log-log axes of panel a) reveal the Basquin-like scaling $\tauf\sim\sig^{-\beta}$ at low $\sig$, as seen experimentally with $\beta=5.5$ in Ref.~\cite{leocmach2014creep} and $\beta=2.0$ in Ref.~\cite{bauland2023non}. This scaling predicts that any stress, however small, will always eventually cause failure. This has  important implications for the shelf life of gel-based products (which are subject to gravity at least), and contrasts notably with the physics of soft glassy materials such as dense emulsions, colloids and microgels in which the timescale of yielding  diverges at a non-zero yield stress $\Sigma_{\rm y}$, with indefinite creep for $\sig<\Sigma_{\rm y}$~\cite{sentjabrskaja2015creep,ballesta2016creep,siebenburger2012creep,divoux2011stress}. The log-lin axes in b) reveal $\tauf\sim\exp(-m\sig)$ at larger $\sig$. 

A notable achievement of the experiments in Ref.~\cite{leocmach2014creep} was to fit the shear rate across all three time regimes -- primary, secondary and tertiary -- to the single master scaling form of Eqn.~\ref{eqn:master}. To explore whether such a fit holds here, we plot the strain rate $\gdot(t)$ in three different ways in Fig.~\ref{fig:creepB}. Panel a) shows the normalised strain rate $\gdot/\gdotmin$ vs normalised time $t/\tauf$ on log-log axes, to emphasise the primary creep regime. The dotted line shows the power $\gdot\sim t^{-\alpha}$, with $\alpha=x$. Panel b) shows the same quantity, now on lin-lin axes to emphasise the secondary crossover regime. As seen experimentally, this secondary regime is seen for times $0.1\lesssim t/\tauf\lesssim 0.9$, with the shear rate attaining a minimum during it at $t=\taumin$. Finally, panel c) plots the normalised shear rate $\gdot/\gdotmin$ vs normalised time to failure, $(\tauf-t)/\tauf$. In this figure, for consistency with the counterpart figure in the experimental reference~\cite{leocmach2014creep}, we have inverted the time axis so as to progress towards failure from left to right. In this way, the limit $\tauf-t\to 0$ at the right hand side of the figure  corresponds to the limit in which $t\to\tauf$ and the sample finally catastrophically fails. The $+1.0$ power seen experimentally is indicated by the dotted line. 
The dashed line in all three panels shows the single experimental master scaling of Eqn.~\ref{eqn:master}. This can be seen to perform well in the primary and tertiary regimes, but significantly less well in the secondary crossover regime.

\begin{figure}[!t]
\includegraphics[width=\columnwidth]{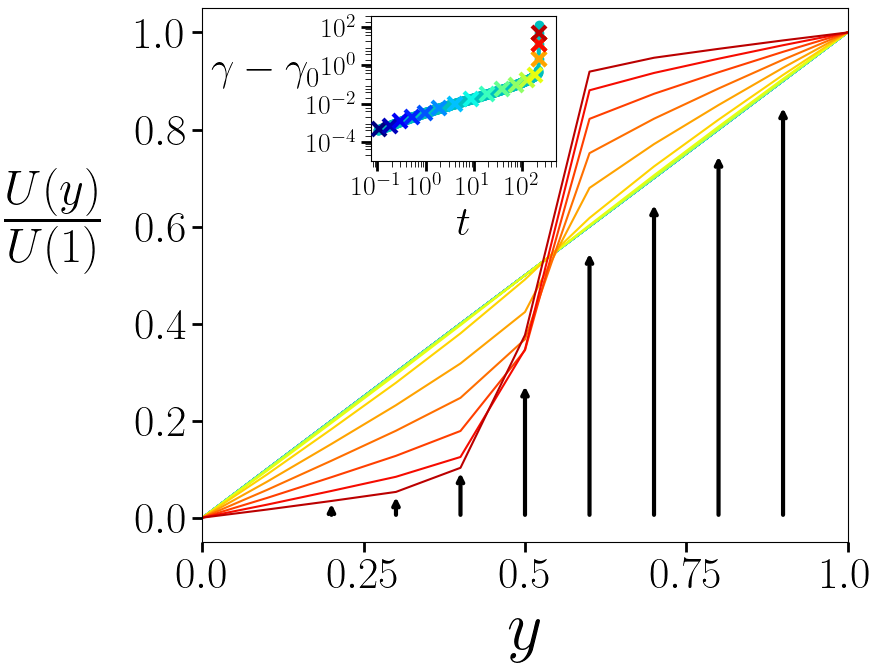}
\caption{Displacement profile $U(y)$ as a function of position across the gradient direction $y$, normalised by the displacement at $y=1.0$. Profiles are shown for times $t$ indicated by symbols in the creep curve $\gamma(t)$ of the inset. The colour of each displacement curve indicates the corresponding time in the creep curve. Imposed stress $\sig=1.6$. Noise temperature $x=0.3$. Waiting time $\tw=10^5$. $\eta=0.05$.
}
\label{fig:displacement}
\end{figure}

Another notable achievement of  Ref.~\cite{leocmach2014creep} was to establish a linear Monkman-Grant relationship between the time $\taumin$ at which the shear rate attains its minimum in the secondary regime, with the later time $\tauf$ of catastrophic material failure in the tertiary regime. Such a relation suggests that the approaching time of material failure might be {\em forecast} in a given experiment, once the earlier shear rate minimum has been observed. Fig.~\ref{fig:MGrelation} shows  the same relation to hold across simulations performed for a wide range of $x, \tw$ and $\sig$. The dashed line shows the linear relationship $\tauf=\taumin/c$ with the same prefactor $c=0.56$ reported experimentally.

Finally in Fig.~\ref{fig:displacement} we plot the displacement profile $U(y)$  at several times during creep and yielding. During early creep, the displacement profile remains essentially linear, corresponding to a near-uniform strain strain profile $\gamma(y)=\partial_y U(y)$, consistent with the lack of strain localisation in the imaging experiments in this regime~\cite{leocmach2014creep}. In contrast, as the material later yields, strong strain localisation arises, as seen experimentally.

\section{Conclusions}

To summarise, we have introduced a mesoscopic model for the irreversible breakup of gels. Its predictions under stress capture the experimentally observed phenomenology, including a primary regime of power-law creep, a secondary cross-over regime in which the shear rate attains a minimum, and a tertiary regime in which the shear rate diverges in a finite time singularity (for $\eta\to 0)$, signifying  material failure. Our results  also confirm  the Basquin-like  and Monkman-Grant scalings of failure time seen experimentally. The model furthermore predicts low levels of material damage during primary creep, later giving way to macroscopic strain localisation and material failure, as seen experimentally. In contrast, theories based on simpler so-called fibre bundle models~\cite{pradhan2010failure}  predict  Andrade creep and failure but only above a critical load~\cite{jagla2011creep}; or Basquin's law but without creep~\cite{kun2008universality,halasz2012competition}.

Despite these successes, our approach has several shortcomings. First, we assumed for simplicity a fast gel strand (re)formation time $\tref = 0.0$ compared to the slow timescale of initial gelation, but an infinite $\tref$ compared to the faster timescale of yielding under stress. Future work might consider a finite $\tref$ consistently during gelation and yielding. Arguably, however, our approach does not accurately model the process of gelation itself: the dynamics for $t<0$ might better be seen as leading to a sensible initial condition for the creep simulations. As noted above, setting $\tref
=0$ throughout, including during straining, recovers the original SGR model. This has been used in earlier work~\cite{moorcroft2013criteria} to model creep and yielding of soft glassy systems, including gels, that flow after yielding. Second, the experiments of Ref.~\cite{leocmach2014creep} suggested part of the primary power law creep to arise from reversible viscoelastic squeezing of fluid through the gel matrix. In contrast, in our model all creep arises via local plasticity. Indeed, the origin of power law creep in amorphous materials is controversial, having been variously attributed to plastic rearrangements~\cite{nechad2005creep} and linear viscoelasticity~\cite{gobeaux2010power,leocmach2014creep}. Finally, we have modelled spatial heterogeneity only in the gradient direction $y$, whereas significant heterogeneity was also observed experimentally in the vorticity direction $z$. Future modelling efforts should increase the dimensionality simulated.

\section*{Conflicts of interest}

There are no conflicts to declare.

\section*{Acknowledgements}
This project has received funding from the European Research Council (ERC) under the European Union's Horizon 2020 research and innovation programme (grant agreement No. 885146).  We thank Andrew Clarke for discussions and SLB (Schlumberger Cambridge Research Ltd.) for support.


%

\end{document}